# Evaluating the Linguistic Coverage of OpenAlex: An Assessment of Metadata Accuracy and Completeness

Lucía Céspedes*[1,2], Diego Kozlowski[1], Carolina Pradier[1], Maxime Holmberg Sainte-Marie[1,3,4], Natsumi Solange Shokida[1], Pierre Benz[1], Constance Poitras[1], Anton Boudreau Ninkov[1], Saeideh Ebrahimy[1,5], Philips Ayeni[1,6], Sarra Filali[1,7], Bing Li[8], Vincent Larivière[1, 2, 7, 9, 10]

[1] Chaire UNESCO sur la science ouverte, École de bibliothéconomie et des sciences de l'information, Université de Montréal, Canada

[2] Consortium Érudit, Université de Montréal, Canada

[3] Danish Centre for Studies in Research and Research Policy, Department of Political Sciences, Aarhus University, Denmark

[4] Department of Design, Media and Educational Science, University of Southern Denmark, Denmark

[5] Department of Knowledge and Information Science, Shiraz University, Iran

[6] Chancellor Paterson Library, Lakehead University, Canada

[7] Centre Interuniversitaire de Recherche sur la Science et la Technologie, Université du Québec à Montréal, Canada

[8] Institute of Science of Science and S&T Management, Dalian University of Technology, China

[9] Department of Science and Innovation, National Research Foundation Centre of Excellence in Scientometrics and Science, Technology and Innovation Policy, Stellenbosch University, South Africa

[10] Observatoire des sciences et des technologies, Université du Québec à Montréal, Canada

* corresponding author: lucia.cespedes@umontreal.ca

## Abstract

Clarivate's Web of Science (WoS) and Elsevier's Scopus have been for decades the main sources of bibliometric information. Although highly curated, these closed, proprietary databases are largely biased towards English-language publications, underestimating the use of other languages in research dissemination. Launched in 2022, OpenAlex promised comprehensive, inclusive, and open-source research information. While already in use by scholars and research institutions, the quality of its metadata is currently still being assessed. This paper contributes to this literature by assessing the completeness and accuracy of OpenAlex's metadata related to language, through a comparison with WoS, as well as an in-depth manual validation of a sample of 6,836 articles. Results show that OpenAlex exhibits a far more balanced linguistic coverage than WoS. However, language metadata is not always accurate, which leads OpenAlex to overestimate the place of English while underestimating that of other languages. If used critically, OpenAlex can provide comprehensive and representative analyses of languages used for scholarly publishing, but more work is needed at infrastructural level to ensure the quality of metadata on language.

**Keywords**

languages in science, bibliometric databases, metadata quality, open access, OpenAlex

**Introduction**

Developed and maintained by the non-profit organization OurResearch, OpenAlex is a freely available, large-scale database of scholarly works. As such, it provides access to a vast multilingual catalog of Works (scholarly documents such as research articles, books, theses, etc.), Authors, Venues (the journals, conferences, repositories, etc., where the Works are hosted), Institutions, Concepts (the disciplines and topics Works are about), and connections between them (citations). Built on an open codebase, its main data sources are the discontinued Microsoft Academic Graph (MAG, of which OpenAlex is presented as a continuation) and Crossref, complemented by an array of smaller-scale databases (ORCID, DOAJ, ROR, Unpaywall, Pubmed and Pubmed Central, the ISSN International Centre, WikiData), disciplinary, national, and institutional repositories (arXiv, Zenodo) as well as fit-for-purpose web crawls (Priem et al., 2022; Piwowar et al., 2022). In addition, the fully open design of OA has the flexibility to potentially integrate other sources.

OpenAlex has received acclaim for its extensive journal and document coverage that facilitates comprehensive analyses. It has been shown to outperform closed, proprietary databases like Clarivate's Web of Science (WoS) and Elsevier's Scopus in terms numbers of documents and journals covered (Alperin et al., 2024; Culbert et al., 2024; van Bellen et al., 2024; Jiao et al., 2023). Among the many advantages associated with the development of a comprehensive database like OpenAlex is the possibility to assess the evolving place of languages in scholarly communication. Recent studies on the topic have either been performed using restrictive databases such as WoS (Larivière, 2018), which, given its indexing criteria, overestimates the place of English, or using surveys, which reduces the generalizability of findings to a place and a point in time (Kulczycki et al., 2018). However, no study has yet assessed the linguistic coverage and data quality of language information in OpenAlex. This is an important gap in the assessment of the platform's performance, since coverage is "a core issue in discussing the data sources that served to globalize biased definitions of excellence, impact, and global standards" (Beigel, 2024, pp.27-28). In this context, language is a highly significant variable, both for bibliometric analyses and for sociological studies of knowledge circulation within and across nations.

One of the main criticisms historically addressed to closed, proprietary databases such as Scopus or WoS is their limited linguistic coverage, where English-language journals are overrepresented and, conversely, journals published in other languages are scarcely indexed. This linguistic bias has been well documented (Ammon, 2006; Archambault et al., 2006; Mongeon & Paul-Hus, 2016; Demeter, 2020), with English documents representing the quasi-totality of Scopus and WoS (Vera-Baceta et al., 2019). The underrepresentation–not to say invisibility–of non-English publications is strongly detrimental to authors and institutions located in non-Anglophone contexts, as the hierarchy and prestige attributed to certain languages of publication, journals, indexes and databases are instruments of gatekeeping and exclusion (Tennant, 2020; Salatino & López Ruiz, 2021; Finardi, 2022; Navarro et al., 2022).

While the centrality and usefulness of English is undeniable, the linguistic landscape portrayed by the oligopoly of commercial publishers does not accurately reflect the linguistic diversity and richness of regional, national, and international scholarly communication networks (Beigel & Digiampetri, 2023; Khanna et al., 2022). Current trends, such as those expressed in the Helsinki Initiative (2019), the UNESCO Recommendation on Open Science (2021), or the Coalition for Advancing Research Assessment (CoARA, 2022), place language diversity and multilingualism in a central role towards building a more open and equitable playing field in international scholarly communications.

In this context, and given its ample and varied sources of information, OpenAlex has the potential to–at least partially–overcome the limitations of previous bibliometric databases in both helping to make visible non-English literature in bibliographic searches, as well as allowing a proper measurement of languages in the dissemination landscape. OpenAlex currently uses the langdetect software library (Danilak, 2021) to infer the language of Works. However, since this algorithm is run only on a work's metadata (abstract, and title if the abstract is missing) and not the full text, language reports are not always accurate, for example, when the language of title and/or abstract is different from the main text, or when metadata is found in more than one language. These limitations are acknowledged by OpenAlex (2024). Thus, if the database is to become a relevant tool in the design and assessment of policies that uphold the value of research conducted in multilingual contexts and published in non-anglophone circuits, a diagnosis of the quality of its linguistic coverage is needed. The completeness and reliability of OpenAlex data must be analyzed both externally, to compare its performance against established benchmarks of bibliometric information, and internally, to assess its consistency. Therefore, in this paper we set out to explore the linguistic coverage of OpenAlex by addressing the following research questions:

**RQ1:** How does linguistic coverage in OpenAlex compare to that of Web of Science?

**RQ2:** How accurate is the *language* label at the article level found in the OpenAlex metadata?

**RQ3:** What are the most common sources of language confusion in OpenAlex?

*State of OpenAlex current and potential uses, strengths and shortcomings*

Given its ample document coverage, open data availability, and good usability through data dumps and REST API, OpenAlex is already being used for bibliometric analysis (Akbaritabar et al., 2023; Jiao et al., 2023; Krause & Mongeon, 2023; Gates et al., 2024; Harder, 2024; Hval et al., 2024; Unzurrunzaga et al., 2024; Vidmar, 2024; to name but a few). OpenAlex has the potential to become a valuable source of information for science policy and research assessment. A growing number of higher education institutions set in contexts as different as France, the Netherlands, Spain, Colombia and Chile, aiming to align themselves with open science practices and principles, have publicly manifested their adoption and support for OpenAlex (Sorbonne Université, 2023; Universidad Central de Chile, 2024; Waltman et al., 2024; CoLaV, 2024; Complexity Lab Barcelona, 2023). Along those lines, the Barcelona Declaration on Open Research Information (2024), which aims to make openness of research information the new norm in order to "advance responsible research assessment and open science and to promote unbiased high-quality decision making", has been signed by more than a hundred organizations that carry out, fund and evaluate research. Likewise, Overton, a platform that indexes policy documents and tracks research papers cited in them, gets author affiliation and documents open access status information from OpenAlex (Overton, 2023a, 2023b).

However, many aspects of OpenAlex's data have been deemed incomplete or inaccurate by the scientometrics community. Scheidsteger & Haunschild (2023) compared OpenAlex's metadata to that of MAG and concluded that OpenAlex seemed to be at least as suited for bibliometric analyses as its predecessor, for publication years before 2021. Alperin et al. (2024) also performed an overall analysis of the suitability of OpenAlex for bibliometric analyses, finding that OpenAlex's coverage outperforms that of Scopus and can be a reliable alternative in many cases. Nevertheless, they also highlighted issues of metadata accuracy and completeness. Likewise, Larivière et al. (forthcoming) showed strong national differences in the increased coverage provided by OpenAlex compared to WoS, with very high increases for the US and Brazil, and much smaller increases for China and many European countries. van Bellen et al. (2024) managed to retrieve twice as many articles from Dimensions and OpenAlex, compared to the more exclusive Web of Science. Therefore, they concluded, if only WoS was considered, it would appear that the "oligopoly" of scholarly publishers persists; however, the inclusiveness of both Dimensions and OpenAlex point at a growing share of smaller publishers.

Analyses into OpenAlex's management of open access information have yielded mixed results. Jahn et al. (2023) revealed inconsistencies in the implementation of open access labeling. Particularly, when performing searches in the database, the authors found that the is_oa filter, which indicates the availability of full texts, did not always match the actual open access status of documents. Simard et al. (2024) concluded that both DOAJ and OpenAlex can be used to compensate for the lack of coverage of diamond journals in mainstream databases, and thus offer greater visibility to such journals. However, the authors observed that the reliance of

OpenAlex on multiple data sources may cause their metadata to be incomplete or inconsistent, in contrast to the more complete and uniform metadata structure of articles published and included in well-resourced journals and indexes.

Comparative analyses with other databases reveal that OpenAlex competes well in terms of reference coverage but falls short in precise classification and accurate metadata management. Culbert et al. (2024) compared OpenAlex's reference coverage and metadata to that of Web of Science and Scopus. For publications included in all three databases, OpenAlex showed average source reference numbers and internal coverage comparable to WoS and Scopus, captured more ORCID identifiers, and a similar quantity of open access status information, but less abstracts and more inconsistencies in its handling of references. Similarly, Delgado-Quirós & Ortega (2024) compared the completeness of publication metadata in eight free-access scholarly databases. Their results showed that Dimensions, OpenAlex, Scilit, and The Lens have higher metadata accuracy and completeness rates than Google Scholar, Microsoft Academic, and Semantic Scholar, but they also signaled the loss of information derived from the practice of integrating data from different sources.

Haupka et al. (2024) analyzed the classification system of OpenAlex alongside those of Semantic Scholar and PubMed, in comparison with Scopus and Web of Science, and determined that OpenAlex and Semantic Scholar lack the detailed categorization found in the other databases. Schares (2024) compared OpenAlex's funder metadata to that of Dimensions', using the United States National Science Foundation as a case study, and found that OpenAlex aggregates funding acknowledgments less precisely than Dimensions, potentially obscuring detailed funder information. Finally, Zhang et al. (2024) identified missing institutional data in OpenAlex as a prevalent problem, especially for articles in earlier years, some research fields like Art and History, and certain journals and publishers.

**Data and Methods**

The main source of information for this project is OpenAlex, which is originally available as JSON objects. We used the Curtin Open Knowledge Initiative (COKI)[1] infrastructure (Hosking et al., 2022), which ingested OpenAlex on a Google Big Query environment, that can be queried via SQL. For comparison, we also used information from Clarivate's Web of Science, which is accessed through the Observatoire des Sciences et des Technologies (OST)[2], that hosts the data as a SQL relational table.

The validation process included three rounds of manual language checking. First, we verified the language of a stratified sample of 50 articles for each of the 55 languages that appear in OpenAlex between 2000 and 2020. Given that some languages did not reach the 50 paper threshold, the first sample accounted for 2,701 articles (instead of 2,750). The second round of manual labeling focused on the 11 languages that represent more than 95% of the articles in OpenAlex as reported by the platform: Chinese, English, French, German, Indonesian, Italian, Japanese, Korean, Portuguese, Russian, and Spanish. We then built a new stratified sample of 285 articles for each of those languages constituted by 15 articles per OpenAlex Concept (a total of 3,135 articles). Moreover, given the number of articles in English, an additional sample of 1,000 English articles was made to ensure a proper representation of the population.

---

[1] https://openknowledge.community/
[2] https://www.ost.uqam.ca/

In order to compare the language assigned by OpenAlex with the actual language of each article, a team of 14 hand coders verified, for all 6,836 articles, the language in which the corresponding full-text was written. Many of the languages in the sample were either spoken or familiar to the members of the team; when this was not the case, online translation tools were applied to portions of the articles' full texts. Full texts were retrieved following the links provided by OpenAlex, or, alternatively, by looking for each article through online university repositories or academic search engines. Corrected languages were coded according to the ISO 639-1 international standardized list. In the case of Chinese, we unified the codes zh_cn (Simplified, People's Republic of China) and zh_tw (Traditional, Taiwan) as zh. For those cases where the full text was not retrievable, we either relied on language metadata provided by reliable sources (such as PubMed, Dialnet for Spanish or CiNii for Japanese), or labeled the observation as Not Found (986 cases) or as No Answer if the language could not be determined (103 cases). The final dataset consisted of 5747 verified articles. Each hand coder also recorded an overall evaluation of the subsamples they had to verify.

Given that we built a stratified sample of articles by language, the expansion factor of each observation results from the ratio between the total number of documents in that language and the number of observations that we were able to label (i.e. where we found the article and were able to assign a language). Using these expansion factors as weights, we computed the precision, recall, and balanced accuracy (i.e. the arithmetic mean between recall and specificity) for each language. Similarly, to compute the corrected frequencies for each language, we extrapolated the information in our confusion matrix and resorted to the ratio between the observed and declared languages to compute the total estimated articles per language.

It is worth mentioning some of the limitations of these methods. For documents in our sample, no relation was found between having an assigned DOI and linguistic classification accuracy. Disciplinary classifications were disregarded as an analytical category, since during manual labeling it became clear that misclassifications were widespread among Concepts in OpenAlex.[3] Besides, while we are aware that OpenAlex employs a broad definition of "article" (including research articles, proceedings, and preprints), we found a significant number of documents that were not research articles, and some were not even documents published in academic journals (1,107 documents, 19%). As regards representativity, it is true that a sample whose language subsample sizes are based on an article language distribution that we expect to be inaccurate is bound to impact the reliability of the results to some extent. However, the language metadata provided by OpenAlex represents the only reliable source of information regarding the language distribution in the database, and thus the only viable basis for any representative sampling strategy for now.

---

[3] At the time of writing this paper, OpenAlex was replacing its original Concepts classification with Topics. Concepts are still provided for Works, but no longer actively maintained nor updated.

**Results**

The language metadata in OpenAlex is consistent in terms of completeness: 98% of the Works indexed provide information on language, while 2% lack this information. We first compared the *declared* distribution of languages in OpenAlex and WoS. Figure 1 presents the percentage of papers by language in OpenAlex and WoS for the 11 languages with the highest proportion of papers in OpenAlex (all other languages are combined in the *Other* category).

The language diversity—in terms of the *declared* language distribution by each platform—is much higher in OpenAlex than in WoS. Specifically, while more than 96% of all papers indexed in WoS are in English, this percentage is 75% in OpenAlex. Striking differences across languages are observed, however, in terms of their relative underrepresentation in WoS. While German and French are 4 times more present in OpenAlex than in WoS, Japanese and Korean papers are respectively found 31 and 60 times more frequently in OpenAlex than in WoS. Indonesian, virtually nonexistent in WoS, accounts for more than 1% of the papers in OpenAlex, surpassing Russian and Italian.

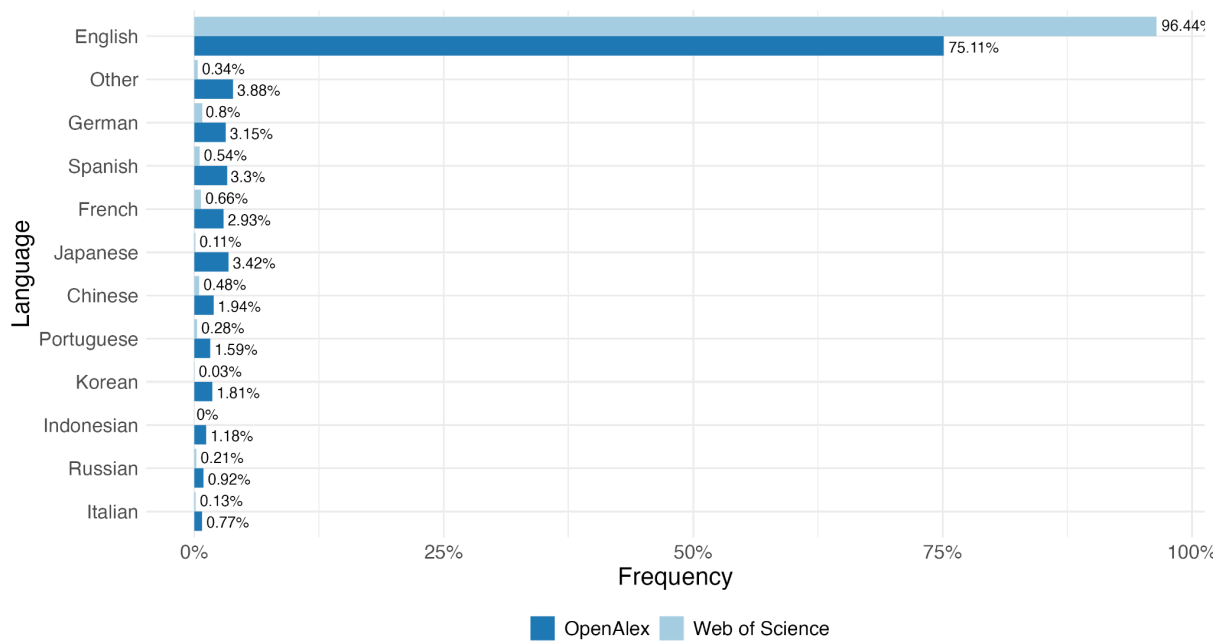

**Figure 1.** Declared percentage of papers by language for OpenAlex and Web of Science, for articles published between 2000 and 2020.

These percentages should not be taken at face value. Our manual validation of the linguistic metadata in our samples found that 14.7% of papers in OpenAlex constitute false positives, this is, they are actually not in the language declared by the platform. Figure 2 presents the multiclass confusion matrix for the 11 main languages in the database, plus all other languages combined. The columns represent the language declared by OpenAlex, while the rows represent the verified language of those same papers. Figure 2A shows the proportion of cases by declared language (vertical normalization), and Figure 2B shows the absolute number of articles corresponding to each combination of verified and declared language in OpenAlex as per our expansion factors.

The main diagonal in Figure 2A shows the percentage of true positives, this is, papers for which the language information found in OpenAlex matched the one found through manual verification. Russian, in the top right corner, was found to be correctly classified most of the time (95.5%). Spanish and Portuguese were correctly labeled more than 90% of the time. For Indonesian, Japanese, English and Korean, correct language attribution ranged from 88.8% to 81.5%, while French, German and Italian exhibited lower percentages (between 75.2% and 73.3%). On the bottom left corner of the figure, Chinese was found to be the most frequently misclassified language (63.7%).

Figure 2B shows the number of papers in terms of their predicted language (according to OpenAlex) and observed language (verified through manual validation). Given the large proportion of English articles in the database, false positives for that language account for much of the difference between predicted and observed frequencies observed in other languages. For example, as more than 3% of articles declared as English were actually in Chinese, we estimate there are around 5.5 million articles in Chinese mistakenly classified as English publications. We also estimate that there are around 4 million papers in Russian that appear to be in English in the database. It is important to note that these figures imply that there are more papers in these languages labeled as English publications than the number of English publications labeled as Chinese or Russian. Korean also shows a large number of publications labeled as English (3.6 million), while in the case of French and German, our verifications suggest that there are more papers in English mistakenly classified as French or German than the other way around. In other words, our findings suggest that linguistic confusion is not symmetrical nor bidirectional.

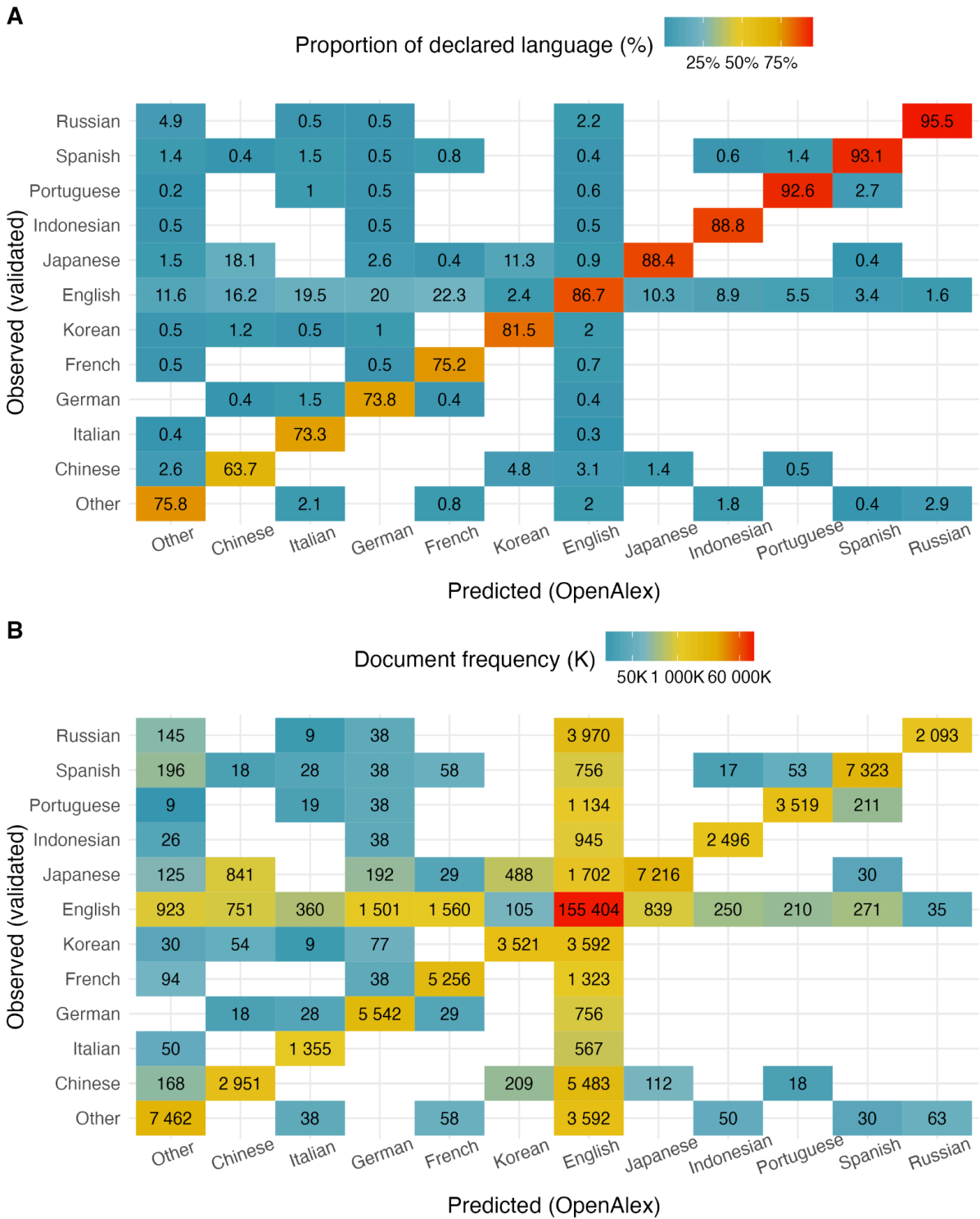

**Figure 2.** Confusion matrix of declared and verified languages in OpenAlex, for a sample of 350 articles per language. The "Other" category corresponds to a sample of 50 articles for each language identified in OpenAlex, totalling 1,631 articles and 43 languages. Articles published between 2000 and 2020.

A glimpse into some evaluation metrics can further complete the picture. As seen in Figure 3, Chinese shows the lowest recall (due to the high number of false negatives, this is, the Chinese articles that were incorrectly classified as English by

OpenAlex), precision, and balanced accuracy. Russian, as verified by our manual coding, appears as the most correctly classified language, however, its low recall indicates the same problem as for Chinese: there is a large share of articles in both languages that OpenAlex misses when queried. Spanish and Portuguese both have strong performances, with Spanish exhibiting the highest recall among the two. Interestingly, Indonesian and Japanese perform in a similar way to the Iberoamerican languages, especially in terms of their balanced accuracy. Italian, French, and German also show similar performances, with high balanced accuracy, moderate to high recall, but moderate to relatively low precision. Finally, while English has a relatively low precision score, its performance in terms of recall is unparalleled amongst all languages (which is expected given its prevalence).

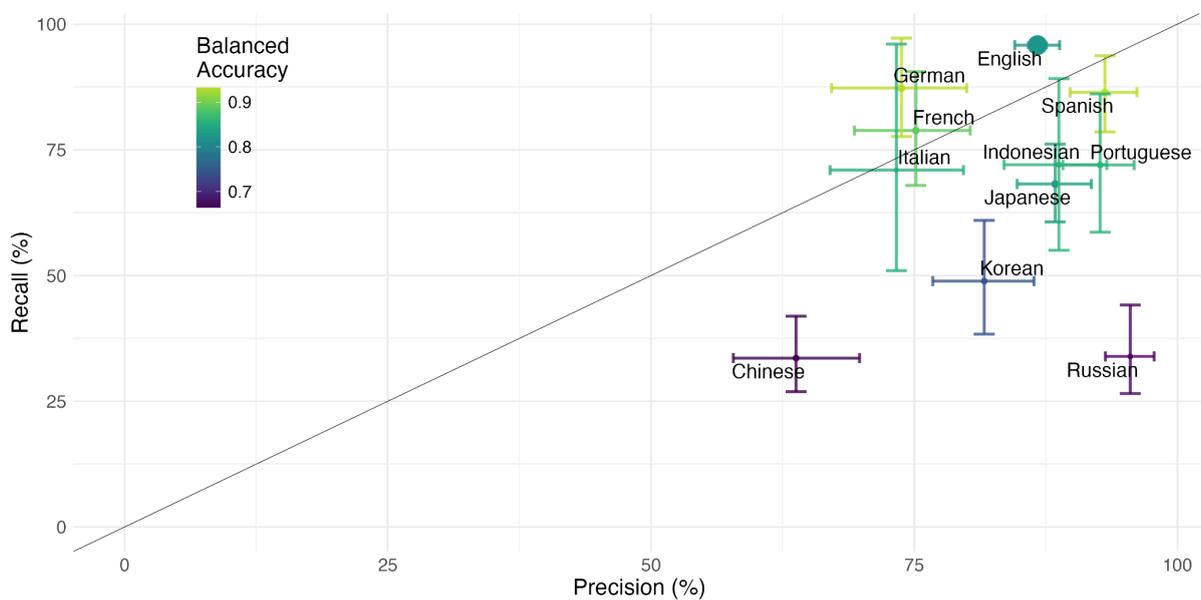

**Figure 3.** Precision and recall of the 11 most common languages in OpenAlex. Balanced accuracy is measured as the average between recall and specificity. Size corresponds to the proportion of articles in the database. Error margins based on 95% stability intervals from 1000 bootstrapped samples. Articles published between 2000 and 2020.

These linguistic misclassifications led us to revise and correct the frequency of each language in OpenAlex. Our data shows that there is even more linguistic variety in OpenAlex than what the database itself provides at first glance. According to our estimations, only 68% of articles are in English, a much lower figure than the 75% declared by OpenAlex or the staggering 96.4% in WoS (Figure 4A). In OpenAlex, English still towers over the rest of the languages, none of which reach even 5%, but, once verified, their frequencies grow in different proportions, revealing a much more linguistically diverse landscape of publications. As seen in figures 4B and 4C, for Japanese, Chinese, Korean, Portuguese, Russian, and the combination of smaller languages ("Other"), the verified proportion is significatively greater than that reported by OpenAlex. The increase of Russian and Chinese from the declared to the corrected frequency is particularly striking, 186% and 93% respectively, meaning there are almost three times more articles in Russian and two times more articles in Chinese than what OpenAlex declares. Spanish, an already very well-classified language, also grows in frequency, though at a lower degree. On the other hand, the actual proportion of French, Italian and, particularly, German papers is lower than reported by OpenAlex.

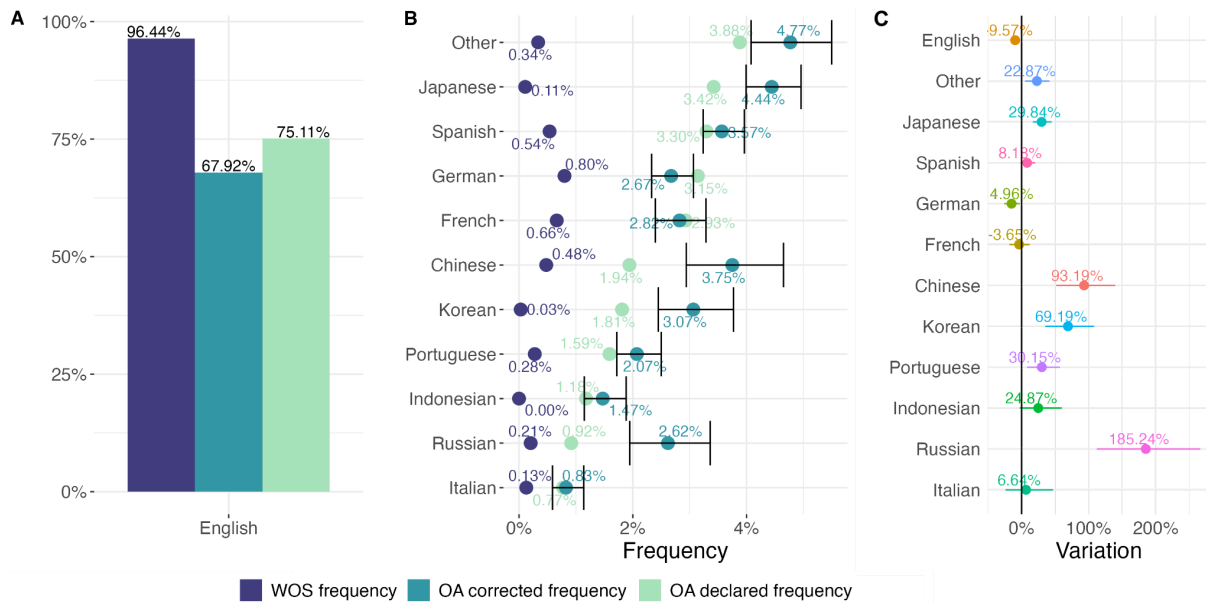

**Figure 4.** OpenAlex declared and corrected frequencies, and WoS frequency of languages, for English (A), and for the following ten most represented languages in OpenAlex (B); and variations between declared and corrected frequencies (C). 95% stability intervals based on 1000 bootstrapped samples.

## Discussion

Our results show that OpenAlex provides a more balanced linguistic coverage than WoS. However, our corrected rates suggest that OpenAlex underestimates its own linguistic diversity, as the relative proportion of indexed documents written in languages other than English is higher than what the database itself suggests. According to the corrected figures, this is, around 68% of articles in English and 32% in all other languages, OpenAlex likely represents a much more accurate reflection of the multilingual nature of scholarly communications worldwide, acknowledging the dominant position of English but also including the languages of science in regional or non-mainstream circuits, and by non-commercial publishers (universities, learned societies, small specialized editors, etc.). This resonates with Khanna et al. (2022), who argue that, contrary to common assumptions, a broader, more linguistically and geographically diverse world of scholarly communication is already underway –but it needs to be visible and properly accounted for. The difference in favor of non-English languages compared to WoS resides in the non-restrictive nature of OpenAlex which, contrary to WoS, indexes documents that have certain metadata and location properties, rather than citation rates.

According to our results, the 11 most prevalent languages in OpenAlex can be grouped in terms of their precision and recall. Our results suggest that OpenAlex is a very reliable source for analysis regarding English and Spanish, since these languages exhibit both high precision and recall.

Indonesian, Japanese and Portuguese show a high precision and a moderate recall. This implies that when retrieving articles in any of these languages, we can be confident of the purity of the data (low false positives), but we cannot guarantee that the whole population will be covered. Consequently, OpenAlex may be confidently used to assess, for example, shared characteristics or topics of scientific publications in any of these languages, or any analysis where the population is that language in itself, but it might arrive at misleading conclusions when comparing those languages with others in terms of relative sizes. On the other hand, French, Italian and German exhibit a moderate precision and moderate to high recall. In these cases, a more comprehensive sample comes at the cost of more statistical noise, particularly for Italian. Therefore, OpenAlex may be used to establish an upper limit for the number of publications in these languages, knowing the figures would likely be an overestimation.

The three remaining languages appear as outliers and cannot be comfortably grouped near others. Russian has the highest precision, but a very low recall. In this case, the same caveats for Indonesian, Japanese and Portuguese apply, and are in fact heightened. Korean stands between moderate and high precision languages, but its recall is considerably lower. Finally, inferences drawn from OpenAlex regarding Chinese should be triangulated and cross-checked with other sources, since the precision and recall values for this language are the lowest of the group.

OpenAlex's reliance on the metadata and not the full text of papers to infer language likely explains its linguistic inaccuracies. Most non-English papers have titles, abstracts, and keywords in English as well, and many English-language papers have paratexts also in other languages; in both cases, it is possible that OpenAlex retrieves the alternative version of the titles and abstracts instead of the language in which the full text is written. Furthermore, the *langdetect* algorithm (Danilak, 2021) only identifies and supports 55 languages, while the ISO 639-1 list includes more than 180. For reference, the Public Knowledge Project registers journals published in 60 languages using their OJS software (Khanna et al., 2022). In the case of our sample, we found that several articles were written in Serbian, a language that is not supported by the algorithm and which is therefore completely omitted. While these biases mostly arise from metadata limitations, algorithmic decisions can also reinforce invisibilization and "may ultimately affect research evaluation and bibliometric results based on their coverage of languages and disciplines" (Simard et al., 2024, p.8).

Our findings also underscore the importance of open interoperable infrastructures for equitable scholarly communication. Some of the best performing languages are those with repositories associated with or directed to specific linguistic communities. Such infrastructures are especially significant for minoritized languages that are

underrepresented–or even excluded–in mainstream scientific information databases, but also for languages with a speaker base and academic communities large enough to establish local or regional circuits of publication with relative degrees of autonomy from English, as Milia et al. (2022) have also argued. For example, the good performance observed for Spanish and Portuguese may be reflecting the strength of the open information infrastructures developed in Iberoamerica. Specifically within the Latin American ecosystem of scientific information databases, we not only find established digital libraries (e.g., Scielo, Redalyc, Latindex), but also a complex network of open access repositories whereby full-texts works by Latin American scholars, a majority of which are in Spanish and Portuguese, are curated and made publicly available. Such infrastructures allow OpenAlex to retrieve high quality metadata even from non-mainstream journals in Spanish and Portuguese. The synergy between open access repositories and good quality in open access bibliometric databases reinforces the importance of the green route to open access[4], which has been showing a decline in the last years as the gold route (nowadays co-opted by APC based open access journals) has been steadily rising (Piwowar et al., 2018; Zhang et al., 2022; Butler et al., 2023).

**Conclusions**

This study shows that OpenAlex's broader coverage translates into a better representation of languages used in scientific communications beyond the mainstream circuits of publication. Our results show that language metadata in OpenAlex offers a representative linguistic portrait of the global scientific communication output, allowing for reliable and thorough linguistic analyses. Our results can orient researchers as regards what questions can be posed for different languages relying on OpenAlex with relatively high confidence. This is particularly relevant in the context of current efforts towards overcoming the limitations of closed, proprietary databases such as Scopus and Web of Science. As Beigel (2024) states, "it is critical to have new open infrastructures that can shed light on bibliodiversity and multilingualism, capable of showing diversified profiles of scientific production and multiscalar research agendas." (pp.27-28). In this line, this first diagnosis of linguistic coverage and accuracy in OpenAlex may be the basis of further studies adopting a sociolinguistic perspective on the publication and circulation of scientific knowledge.

Given the completeness of OpenAlex's language data, efforts should be oriented to verifying and correcting the quality and accuracy of its linguistic metadata. Indeed, metadata quality is crucial for the retrieval of these documents, but also for the measurement of various aspects of scholarly communication, such as language. Therefore, all scientific information and dissemination infrastructures, including but not limited to those which OpenAlex relies on for its metadata, should be "empowered" to be able to have better metadata quality. Thematic, institutional and national repositories are heterogeneous and not all of them show the same granularity of metadata, especially in terms of language, nor are they integrated in a systematic, coordinated fashion. Thus, improving metadata quality at the sources would reduce the potential for carry-over errors from the original documents to their publication venues to the platforms that databases like OpenAlex harvest. Collective curation of small subsets of OpenAlex, such as the one conducted for this study, may also be undertaken as a long-term initiative. The relationship between open access, availability of metadata, and language accuracy should be further studied. Both disciplinary classification and work type are also potential areas for future research and improvement within the platform (see Haupka et al., 2024, for a first insight into the publication and document types in OpenAlex).

---

[4] Green open access refers to the practice of self-archiving in institutional, thematic or national public repositories. Gold open access entails publishing in journals that make articles immediately free and accessible to readers. If these journals do not charge any sort of fee for authors either, they are considered diamond open access. Finally, bronze open access comprises articles free for readers but without any identifiable license.

OpenAlex is a highly dynamic database, and it is to be expected that metadata quality and coverage will evolve as a user community consolidates around it. As more studies identify its strengths and limitations, and pinpoint specific areas for improvement, OpenAlex will be able to benefit from this user-based feedback. This synergy is characteristic of open science and open innovation initiatives and will be key to the development and establishment of OpenAlex as a trusted source in the scientific information landscape.

**Acknowledgements**

The authors thank Aditi Ashok for her valuable help with the manual verification of languages.

**Funding**

Lucia Céspedes, Diego Kozlowski and Vincent Larivière acknowledge funding from the Social Science and Humanities Research Council of Canada Pan-Canadian Knowledge Access Initiative Grant (Grant 1007-2023-0001), and the Fonds de recherche du Québec - Société et Culture through the Programme d'appui aux Chaires UNESCO (Grant 338828). Pierre Benz acknowledges funding from the Swiss National Science Foundation in the framework of its Postdoc.Mobility scheme (grant number: 210805).

**Competing interests**

None of the authors have conflicts of interest to declare.

**Author contributions**

| CRediT task | Author initials |
| --- | --- |
| Conceptualization | LC, DK, VL |

| Data curation | LC, DK, CPr, MHSM, NSS, PB, CPo, ABN, SE, PA, SF, BL, VL |
|---|---|
| Formal analysis | DK, CPr |
| Funding acquisition | VL |
| Investigation | LC |
| Methodology | DK, CPr |
| Project administration | LC, VL |
| Resources | VL |
| Supervision | LC, VL |
| Visualization | DK, CPr |
| Writing – original draft | LC, DK, VL |
| Writing – review & editing | LC, DK, CPr, MHSM, NSS, PB, CPo, ABN, SE, PA, SF, BL, VL |